\documentclass{IEEEtran}
\usepackage[dvipsnames]{xcolor}
\usepackage{cite}
\usepackage{amsmath,amssymb,amsfonts}
\usepackage{mathtools}
\usepackage{lipsum}
\usepackage{algorithmic}
\usepackage{graphicx}
\usepackage{textcomp}
\usepackage{booktabs}
\usepackage{esint}
\usepackage{xcolor}
\def\BibTeX{{\rm B\kern-.05em{\sc i\kern-.025em b}\kern-.08em
    T\kern-.1667em\lower.7ex\hbox{E}\kern-.125emX}}
\begin{document}
\title{Design of Conformal Array of Rectangular Waveguide-fed Metasurfaces}
\author{Insang Yoo, \IEEEmembership{Student Member, IEEE}, and David R. Smith \IEEEmembership{Senior Member, IEEE}
\thanks{The authors are with the Department of Electrical and Computer Engineering, Duke University, Durham, NC, 27708 USA (e-mail: insangyoo1@gmail.com).}}

\maketitle

\begin{abstract}
We present a systematic design method for a cylindrical conformal array of rectangular waveguide-fed metasurfaces. The conformal metasurface consists of multiple curved rectangular waveguides loaded with metamaterial elements\textemdash electrically small irises\textemdash inserted into the upper conducting walls of the waveguides. Each element radiates energy into free space to contribute to an overall radiation pattern. Thus, the geometry or electrical configuration of each of the individual metamaterial elements needs to be tailored to generate a desired pattern. In general, due to difficulties in modeling the effect of curvature, the design of conformal metasurface arrays has relied on full-wave simulations or experiments. In this study, we propose a design method utilizing the analytic model of a planar metasurface accounting for metamaterial elements' locations and orientations over a surface with curvature. Although approximate, we demonstrate that such an alteration along with the framework of dipolar modeling of planar elements can be used for the analysis of conformal arrays with small curvature. We then design a conformal array metasurface using the method combined with CMA-ES optimizer. Through numerical simulations, we confirm the validity of the proposed design method. Applications include the design of metasurfaces for radar, communications, and imaging systems for automobiles and airplanes.
\end{abstract}


\section{Introduction} \label{sec:introduction}

\IEEEPARstart{C}{}onformal antennas are essential components in applications that require the integration of radiating structures with curved structures, as in many mobile platforms \cite{josefsson2006conformal}. As highly directive beams are generally required for conformal antennas, phased arrays with curved patches \cite{kashiwa1994analysis,loffler1997conformal} or planar patches on multifaceted surfaces \cite{schippers2008conformal,sun2010application} have been used for their beamforming capability via controlling the phase of each radiating element. However, conformal phased arrays often suffer from the increased complexity in their feed system and usage of costly components such as phase shifters. To mitigate these problems, conformal leaky-wave antennas have been suggested as a solution that can offer high directivity and simple feed network \cite{ohtera1990focusing,ohtera1999diverging}. Accordingly, various types of conformal leaky-wave structures have been proposed to generate narrow beams and implement beam scanning capability using tapered microstrip lines \cite{radcliffe2006microstrip,radcliffe2007finite}, substrate integrated waveguides \cite{martinez2014conformal,bayraktar2014circumferential}, and metamaterial transmission-lines \cite{hashemi2008dispersion,hashemi2009electronically}. Despite the advantages offered by the leaky-wave antennas, they rely on specific leaky modes for radiation, which necessitates gradual change or periodicity in their geometry \cite{oliner2007leaky,jackson2012leaky}. Such an underlying mechanism of the leaky-wave structures may limit design flexibility for satisfying system constraints of applications at hand.

Recently, waveguide-fed metasurfaces have emerged as a low-cost, light-weight platform that can provide nearly full control over shaping radiation patterns \cite{smith2017analysis}. In a waveguide-fed metasurface antenna, metamaterial elements\textemdash electrically small, complementary electric-LC (cELC) resonators \cite{yoo2016efficient}\textemdash are patterned into the conducting walls of a waveguide and are used as radiators that are fed directly by guided modes, thus avoiding a complex feed network. The ample control over the radiation patterns formed by a metasurface antenna results from a combination of the phase accumulation of the guided modes and the phase shift provided by the resonant elements, each characterized by a Lorentzian polarizability \cite{smith2017analysis,pulido2017polarizability}. Unlike conventional leaky-wave structures, waveguide-fed metasurfaces do not require specific modes for radiation. Hence, the geometry change of metamaterial elements does not need to be gradual, and an aperiodic arrangement of metamaterial elements is allowed to offer greater design flexibility \cite{pulido2016dda,pulido2017discrete,pulido2018analytical,yoo2019analytic,yoo2020full}. Thus, waveguide-fed metasurface antennas have been employed in a variety of applications including computational imaging \cite{hunt2013metamaterial,sleasman2016microwave}, wireless power transfer \cite{smith2017wpt,gowda2018focusing}, synthetic aperture radar \cite{boyarsky2017synthetic}, and wireless communication systems \cite{yoo2018enhancing,yoo2020dynamic}.

\begin{figure}[!t]
\centering
\includegraphics[width=2.0in]{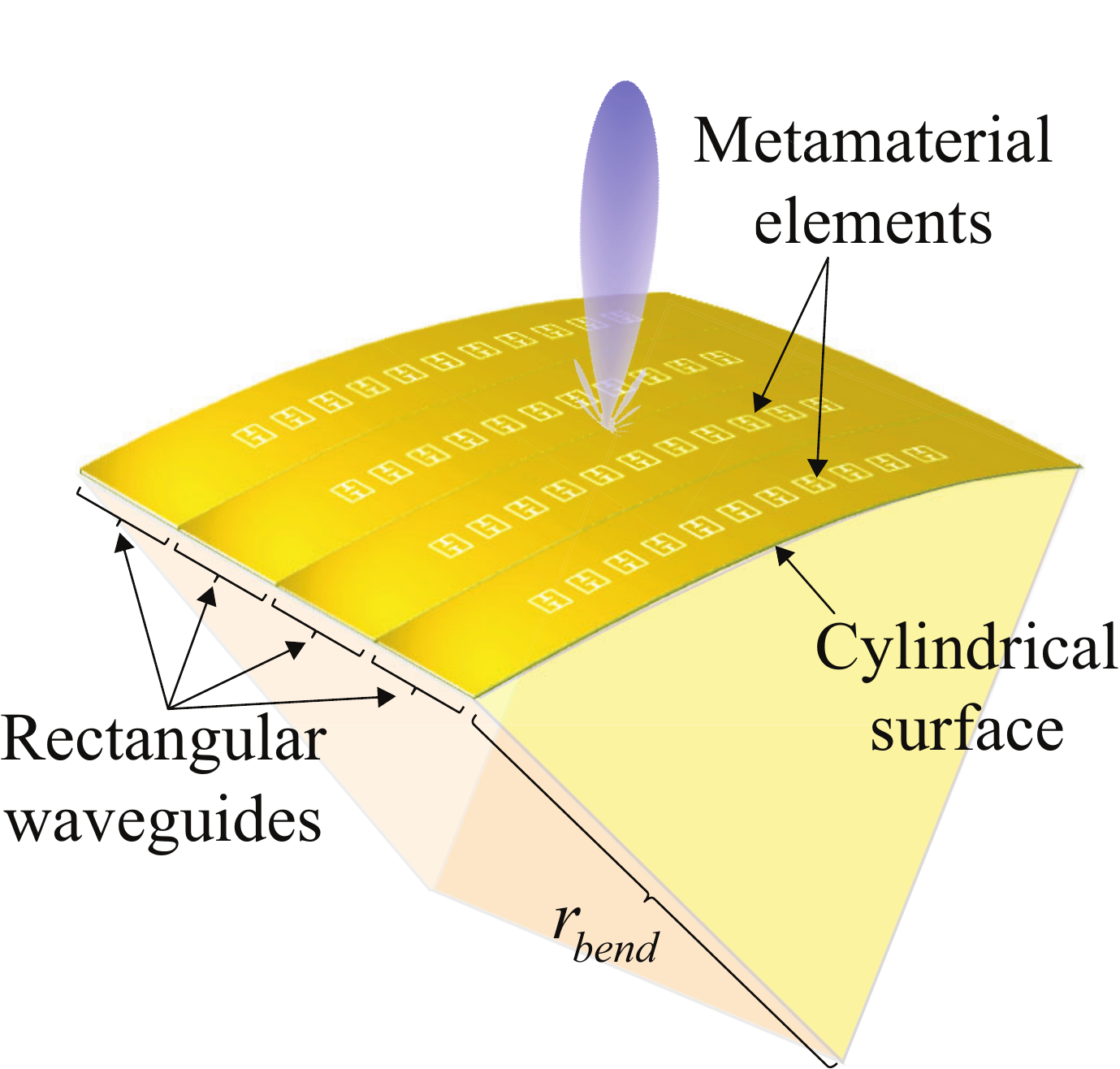}
\caption{Schematic of a cylindrical conformal metasurface antenna consisting of an array of rectangular waveguides embedded with metamaterial elements. The metamaterial elements are electrically small apertures that are etched into the top plates of rectangular waveguides.}
\label{Fig1_SchematicConformal}
\end{figure}

Given the numerous advantages associated with waveguide-fed metasurfaces, their application to conformal antenna systems is of great interest. Among the possible metasurface configurations, printed waveguide-fed metasurfaces\textemdash which we refer to as metasurfaces backed by electrically thin waveguides\textemdash are well-suited for conformal antenna systems owing to their physical nature; printed metasurfaces are lightweight and can be conveniently mounted on curved surfaces. For cylindrical conformal radiating systems, printed rectangular waveguide-fed metasurfaces are promising as the guided modes are relatively unchanged for a large radius of curvature relative to the flat waveguide \cite{lewin1955propagation}, significantly simplifying the design process of conformal metasurfaces. Furthermore, it is straightforward to implement a conformal array by arranging many curved rectangular waveguide-fed metasurfaces together, as illustrated in Fig. \ref{Fig1_SchematicConformal}. As recently demonstrated in its planar counterpart \cite{boyarsky2021electronically}, such an array of rectangular waveguide-fed metasurfaces is attractive because the array system can offer beamforming and beam-steering capabilities using low-power switchable components (e.g., liquid crystal, varactor diodes). The feed network of the array system can be formed simply by using a power-dividing structure. In spite of these merits, up to now, the design and analysis of a conformal array of the printed, rectangular waveguide-fed metasurfaces have not been reported. The reason may be due to difficulties in modeling curved metamaterial elements and their interaction with waveguides.

For planar metamaterial elements, the framework of dipolar modeling of the elements\textemdash which is based on the analysis of the electric field on the planar aperture of a small iris\textemdash is well studied in literature \cite{pulido2017polarizability,yoo2020full,yoo2019model}. Accordingly, a planar metamaterial element with an arbitrary shape can be described as the combination of point electric and magnetic dipoles, so that coupled-dipole methods naturally capture the interactions among the elements \cite{pulido2018analytical,yoo2019analytic,yoo2020full}. For curved metamaterial elements, however, the electric field at one location on the curved aperture is not always tangential to the element's surface, which does not satisfy the assumption used in the dipolar modeling approach. Therefore, the direct application of the dipolar modeling approach to the curved elements is not straightforward. One can assume the field formed on the aperture to model an element such as curved slots \cite{harrington1961timeharmonic}, but the closed-form expression of the field may not be attainable for complex-shaped metamaterial elements.

In this paper, we demonstrate that the framework of dipolar modeling of metamaterial elements along with the analytic model of rectangular waveguide-fed metasurfaces can be used to design cylindrical conformal arrays with a large bend radius. In this approach, we assume that each of the waveguides in the conformal array is piece-wise linear along the circumference of the cylindrical surface. While approximate, we find that the proposed approach not only provides a reasonably accurate description of the conformal array, but also simplifies the design procedure significantly. With the aim of demonstrating the benefits of using the analytic model for the design of the conformal array, design and analysis of a conformal array of the metasurfaces are provided.

\section{Theory} \label{sec:theory}


\subsection{Analytic Model of an Array of Rectangular Waveguide-fed Metasurfaces} \label{sec:theory_part1}

In this subsection, we propose an analytic model of a two-dimensional, planar array of rectangular waveguide-fed metasurfaces by extending the model of a single rectangular waveguide-fed metasurface, reported in \cite{pulido2016dda,pulido2017discrete,pulido2017polarizability}. In the array configuration, rectangular waveguides loaded with metamaterial elements are located in parallel, and the spacing between adjacent waveguides is $s_{wg}$. Each metasurface in the array consists of a dielectric-filled rectangular waveguide with an array of $U$ metamaterial elements embedded in the top wall of the waveguide having width of $a$ and height of $b$. Throughout this work, we assume that the dimensions of the waveguide (i.e., $a$ and $b$) and dielectric constant of the substrate filling the waveguide are chosen such that the dominant mode (i.e., TE$_{10}$) is the only propagating mode. The metamaterial elements are electrically small apertures, each of which can be approximated as point magnetic and electric dipoles, according to Bethe's theory \cite{bethe1944theory}. The dipoles can be characterized by the magnetic and electric polarizabilities that relate the dipole moments to the incident fields \cite{pulido2017polarizability,pulido2018analytical}, given by
\begin{equation} \label{eff_polarizabilities}
\begin{aligned}
\alpha_{xx}^{mm} = m_{x}/H^{inc}_{x}, \
\alpha_{zz}^{mm} = m_{z}/H^{inc}_{z}, \
\alpha_{yy}^{ee} = p_{y}/\epsilon E^{inc}_{y},
\end{aligned}
\end{equation}
with $m_{x}$, $m_{z}$, $p_{y}$ being respectively the magnetic and electric dipole moments representing the metamaterial element. The matrices $H^{inc}_{x}$, $H^{inc}_{z}$ and $E^{inc}_{y}$ represent the incident magnetic and electric fields measured at the center of the element.


Under these assumptions, an array of the rectangular waveguide-fed metasurfaces can be modeled using a coupled matrix equation, given by
\begin{equation} \label{DDA_array}
\begin{aligned}
\begin{bmatrix*}[c]
\mathbf{G}^{}_{11} & \mathbf{G}^{fs}_{12} & \cdots & \mathbf{G}^{fs}_{1K} \\
\mathbf{G}^{fs}_{21} & \mathbf{G}^{}_{22} & \cdots & \mathbf{G}^{fs}_{2K} \\
\vdots & \vdots & \ddots & \vdots \\
\mathbf{G}^{fs}_{K1} & \mathbf{G}^{fs}_{K2} & \cdots & \mathbf{G}^{}_{KK}
\end{bmatrix*}
\begin{bmatrix*}[c]
\mathbf{J}_{1} \\ 
\mathbf{J}_{2} \\
\vdots \\
\mathbf{J}_{K}
\end{bmatrix*}=\begin{bmatrix*}[c]
\mathbf{F}^{inc}_{1} \\ 
\mathbf{F}^{inc}_{2} \\ 
\vdots \\
\mathbf{F}^{inc}_{K} 
\end{bmatrix*},
\end{aligned}
\end{equation}
where $\mathbf{G}^{}_{kk}\in\mathbb{C}^{3U \times 3U}$ is given in \cite{pulido2017discrete,pulido2018analytical} as
\begin{equation} \label{DDA_eq}
\begin{aligned}
\mathbf{G}_{kk} &=
\begin{bmatrix*}[l]
\mathbf{G}^{mm}_{xx,k} & \mathbf{G}^{mm}_{xz,k} & \mathbf{G}^{me}_{xy,k} \\
\mathbf{G}^{mm}_{zx,k} & \mathbf{G}^{mm}_{zz,k} & \mathbf{G}^{me}_{zy,k} \\
\mathbf{G}^{em}_{yx,k} & \mathbf{G}^{em}_{yz,k} & \mathbf{G}^{ee}_{yy,k}
\end{bmatrix*},
\end{aligned}
\end{equation}
and
\begin{equation} \label{DDA_eq_rest}
\begin{aligned}
\mathbf{J}^{}_{k} &=
\begin{bmatrix*}[l]
\mathbf{m}_{x,k} & \mathbf{m}_{z,k} & \mathbf{p}_{y,k}
\end{bmatrix*}^{T},\\
\mathbf{F}^{inc}_{k} &=
\begin{bmatrix*}[l]
\mathbf{H}^{inc}_{x,k} & \mathbf{H}^{inc}_{z,k} & \mathbf{E}^{inc}_{y,k}
\end{bmatrix*}^{T},
\end{aligned}
\end{equation}
where $(\cdot)^{T}$ represents the transpose operator. $\mathbf{m}_{x,k},\mathbf{m}_{z,k}\in\mathbb{C}^{U \times 2}$ and $\mathbf{p}_{y,k}\in\mathbb{C}^{U \times 2}$ are effective magnetic and electric dipole moments representing metamterial elements in $k$th metasurface. The subscripts $``x"$, $``y"$ and $``z"$ refer to $\hat{x}$, $\hat{y}$ and $\hat{z}$ component, respectively. $\mathbf{H}^{inc}_{x,k}, \mathbf{H}^{inc}_{z,k}, \mathbf{E}^{inc}_{y,k}\in\mathbb{C}^{U \times 2}$ represent the magnetic and electric field incident on metamaterial elements in $k$th waveguide, respectively. Note that each column of $\mathbf{F}^{inc}_{k}$ represents the amplitude of the incident fields when the port on one side of $k$th waveguide is excited.

The diagonal entries of $\mathbf{G}_{xx,k}^{mm}$, $\mathbf{G}_{zz,k}^{mm}$ and $\mathbf{G}_{yy,k}^{ee}$ matrices in (\ref{DDA_eq}) are the inverse of the effective polarizabilities of metamaterial elements given in (\ref{eff_polarizabilities}), and their off-diagonal elements are the sum of Green's functions of point magnetic and electric sources in a rectangular waveguide and their free space Green's functions \cite{collin1960field}. In this manner, the off-diagonal entries of $\mathbf{G}_{xx,k}^{mm}$, $\mathbf{G}_{zz,k}^{mm}$ and $\mathbf{G}_{yy,k}^{ee}$ model mutual coupling of metamaterial elements in $k$th metasurface through the scattered fields into the waveguide and free space. The diagonal elements of off-diagonal blocks in $\mathbf{G}_{kk}$ of (\ref{DDA_array}) are zero, while their off-diagonal elements are the sum of the free space Green's functions and those in the waveguide. Note that the matrices $\mathbf{G}^{fs}_{kl}\in\mathbb{C}^{3U \times 3U}$ in (\ref{DDA_array}) capture the coupling of metamaterial elements in $k$th and $l$th metasurface through their radiated fields.

Given the amplitudes of the incident fields (i.e., $\mathbf{F}^{inc}_{k}$), the matrix equation in (\ref{DDA_array}) can be solved for the dipole moments representing all of the metamaterial elements in the array system (i.e., $\mathbf{J}$ matrices). The obtained dipole moments can be used to compute the overall radiation pattern by summing the radiated field by each element. In this manner, the analysis of the metasurface antenna is reduced to solving a matrix equation with size on the order of $(3U\times K)$. It should be noted in (\ref{DDA_eq}) that the field scattered by each metamaterial element can be decomposed into the fundamental mode and many higher-order modes \cite{collin1960field}. Throughout this work, we assume that metamaterial elements are placed at suitable distances, where the first 5 TE modes are sufficient to model the scattered fields by the elements (see Appendix A).

\subsection{Modification of the Analytic Model for Conformal Metasurfaces} \label{sec:theory_part2}

\begin{figure}[!t]
\centering
\includegraphics[width=3.3in]{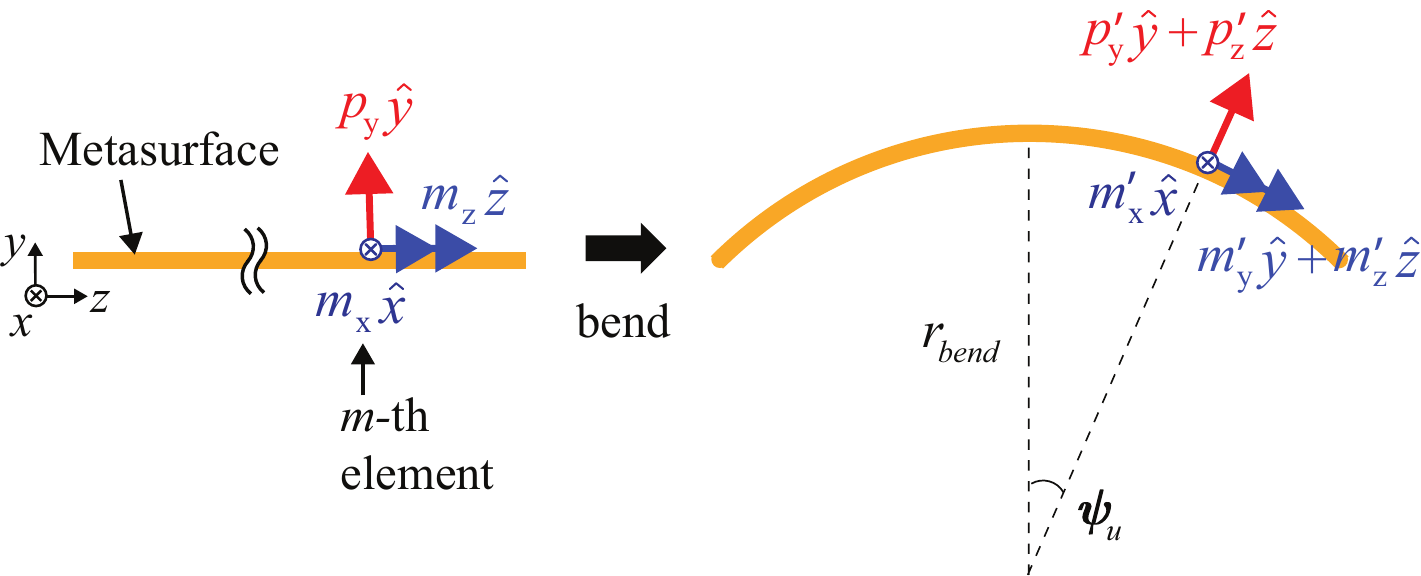}
\caption{Schematic of a planar and cylindrical conformal waveguide-fed metasurface antennas (side view). The metasurface wraps a cylindrical surface (with a radius of $r_{bend}$) along the circumferential direction. Each of the metamaterial elements is represented using the angle $\psi_{u}$, which is measured from $\hat{x}$ axis.}
\label{Fig2_Conformal}
\end{figure}

In this subsection, we modify the proposed analytic model of two-dimensional array of planar rectangular waveguide-fed metasurfaces introduced in Section \ref{sec:theory_part1} for the design of curved metasurfaces. Toward that end, we begin by modeling curved metamaterial elements illustrated in Fig. \ref{Fig2_Conformal}. As discussed, a planar metamaterial element is modeled as a point electric dipole along $\hat{y}$ and magnetic dipoles along $\hat{x}$ and $\hat{z}$. We then bend the metasurface to wrap the surface of a cylinder with a radius $r_{bend}$ along the circumferential direction. Assuming the bend radius is large, we suggest that the cylindrical surface can be considered locally piece-wise linear, such that the electric dipole, located at the center of the element, is in the direction normal to the circumference\textemdash which may be a valid assumption if metamaterial elements are electrically small. Then, the electric dipole moment representing the element can be decomposed into $\hat{y}$ and $\hat{z}$ components, i.e., $p'_y$ and $p'_z$. Similarly, the magnetic dipole moments are parallel to the tangent plane at the center of the element, and accordingly, each $m_{z}$ can be decomposed into $m'_y$ and $m'_z$. Under these assumptions, the the dipole moments in the planar metasurface can be transformed into the dipole moments in the curved metasurface using the following relation, given by
\begin{equation} \label{dipole_conformal}
\begin{aligned}
m'_{x} = m_{x},& \
m'_{y} = -m_{z}\sin\left(\psi_{u}\right), \
m'_{z} = m_{z}\cos\left(\psi_{u}\right), \\
p'_{y} &= p_{y}\cos\left(\psi_{u}\right), \
p'_{z} = p_{y}\sin\left(\psi_{u}\right),
\end{aligned}
\end{equation}
where $\psi_{u}$ represents the angle between the positive $\hat{y}$ axis and the radial direction the metamaterial element from the axis of the cylindrical surface. Note that the proposed modeling approach is approximate and somewhat oversimplified. However, we find that the approach provides reasonably good description of metamaterial elements etched into the cylindrical surfaces.

Once we model metamaterial elements on cylindrical surfaces, the propagation of guided modes in curved rectangular waveguides needs to be examined. Under the assumption that the curvature of the cylindrical surface is sufficiently small, we may further assume that the guided modes in the curved waveguides are similar to those in the planar waveguides with the same cross-section \cite{lewin1955propagation,cochran1966mode}. Accordingly, we suggest that the incident electric and magnetic fields and the Green's functions for a planar waveguide may be used to model conformal metasurfaces with large bend radius.

\section{Design of Cylindrical Conformal Metasurface Antenna} \label{sec:design_ex}

\begin{figure}[!t]
\centering
\includegraphics[width=3.5in]{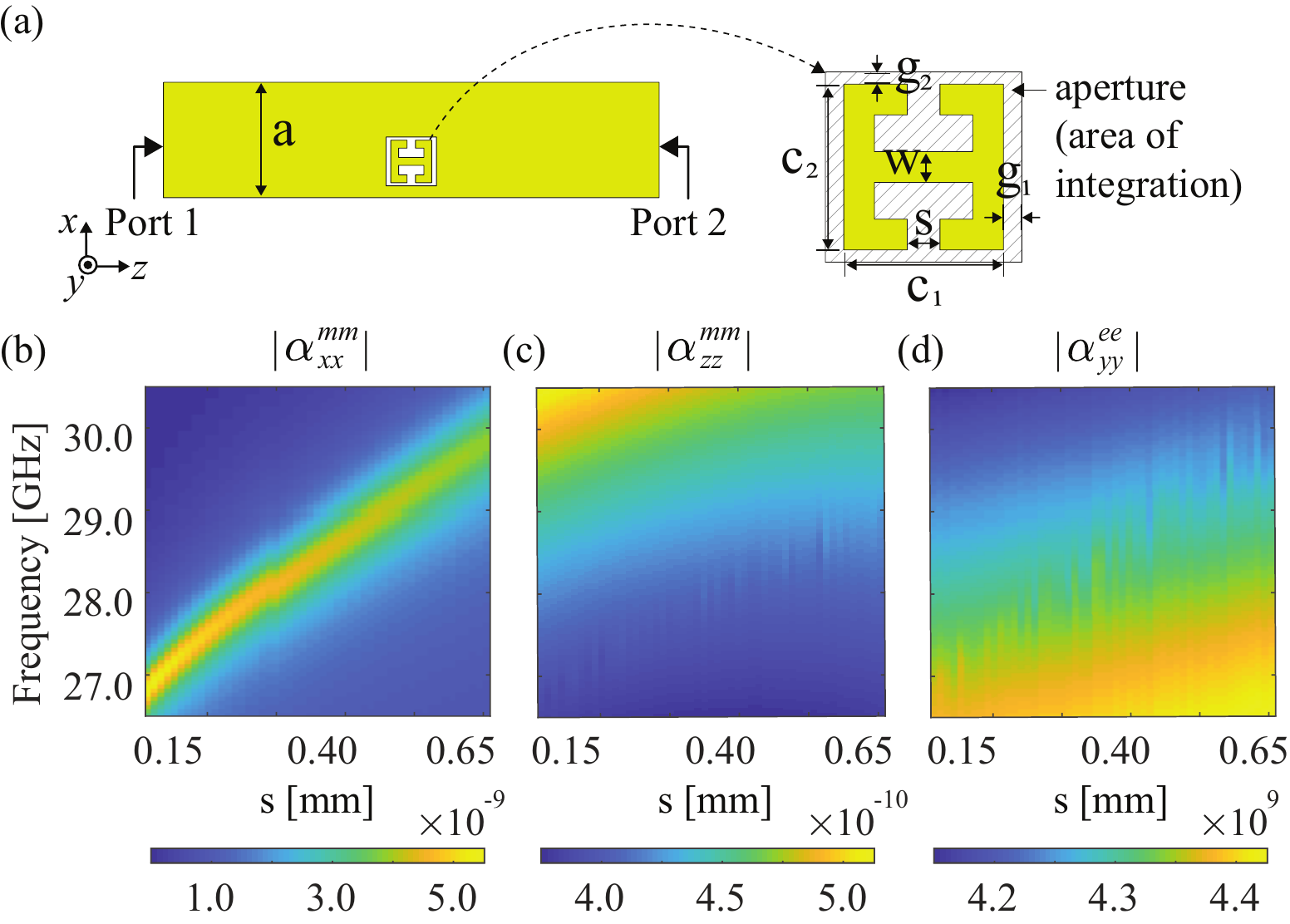}
\caption{(a) Simulation setup for the polarizability retrieval. The width of the waveguide is $a=3.41$ mm. Orange color indicates perfect electric conductor. Design parameters of the metamaterial elements are $c_{1}=1.6$ mm, $c_{2}=1.55$ mm, $g_{1}=0.15$ mm, $g_{2}=0.1$ mm, and $w=0.25$ mm. The amplitude of the polarizability components as functions of frequency and state $s$: (b) $\alpha_{xx}^{mm}$, (c) $\alpha_{zz}^{mm}$, (d) $\alpha_{yy}^{ee}$.}
\label{Fig3_MetamaterialDesign}
\end{figure}

We begin this section with the design of metamaterial elements. The geometry of the element, with key dimensions labeled, is shown in Fig. \ref{Fig3_MetamaterialDesign}(a). The metamaterial element is embedded in a top wall of a planar rectangular waveguide filled with a 0.76-mm-thick Rogers 4003 circuit board. The location of the element is $x=a/5, z=0$. We analyzed the structure depicted in Fig. \ref{Fig3_MetamaterialDesign}(a) using the full-wave electromagnetic solver and computed the dipole moments of the element using the followings, given in \cite{collin1960field,pulido2017polarizability} as
\begin{equation} \label{dipole_moments_formula}
\begin{aligned}
\bar{p} &= \frac{\epsilon_{0}\hat{y}}{2}\iint \left(E_{x}^{a}+E_{z}^{a}\right) dxdz, \\
\bar{m} &= \frac{1}{j\omega\mu_0}\left(-\hat{x}\iint E_{z}^{a} dxdz + \hat{z}\iint E_{x}^{a} dxdz\right),
\end{aligned}
\end{equation}
where $E_{x}^{a}$ and $E_{z}^{a}$ represent the electric fields on the aperture of the element. After that, we computed the effective polarizabilities of the element using (\ref{eff_polarizabilities}). Note that the accuracy of this retrieval method depends on the quality of mesh in the full-wave analysis \cite{pulido2017polarizability}. Thus, we used one-tenths of the shortest length of the element as the maximum length of the mesh. However, we found that the obtained polarizabilities resulted in errors in the predicted S-parameter of the structure in Fig. \ref{Fig3_MetamaterialDesign}(a). Accordingly, the obtained polarizabilities using (\ref{eff_polarizabilities}) and (\ref{dipole_moments_formula}) were multiplied by a real number such that the lease-square error between predicted and calculated S-parameter is minimized.

To find a set of accessible polarizabilities for metasurface design, we swept the gap size $s$ of the metamaterial element from $0.15$ mm to $0.65$ mm and applied (\ref{dipole_moments_formula}) to each element. Figures \ref{Fig3_MetamaterialDesign}(b)-(d) show the retrieved polarizability components as functions of frequency and the gap size $s$. As shown in Fig. \ref{Fig3_MetamaterialDesign}(a), $\alpha_{xx}^{mm}$ exhibits the Lorentzian response for fixed $s$, which can be utilized for pattern synthesis of metausurface antennas \cite{smith2017analysis}.


Having obtained the available polarizabilities of the metamaterial elements, we design a conformal array of the metasurfaces utilizing the analytic model of planar metasurfaces with modifications on the locations and orientations of the dipoles representing the metamaterial elements. The design process starts with the geometric configurations such as the number of waveguides, metamaterial elements, and their locations. We then solve for the dipole moments representing the elements using (\ref{DDA_array}). The locations and computed dipole moments are then modified, according to (\ref{dipole_conformal}) and Fig. \ref{Fig2_Conformal}, to compute the radiation pattern. In the design process, we find the optimal set of polarizabilities of the metamaterial elements\textemdash among the available polarizabilities shown in Fig. \ref{Fig3_MetamaterialDesign}(b)--(d)\textemdash using CMA-ES \cite{hansen2001completely}, which has been reported to work effectively for the design of metasurfaces \cite{gregory2011fast,jiang2014broadband}. While the approach is approximate and simplified, we demonstrate that radiation patterns predicted by the method are reasonably accurate in comparison with full-wave numerical simulations.

\begin{figure}[!t]
\centering
\includegraphics[width=3.5in]{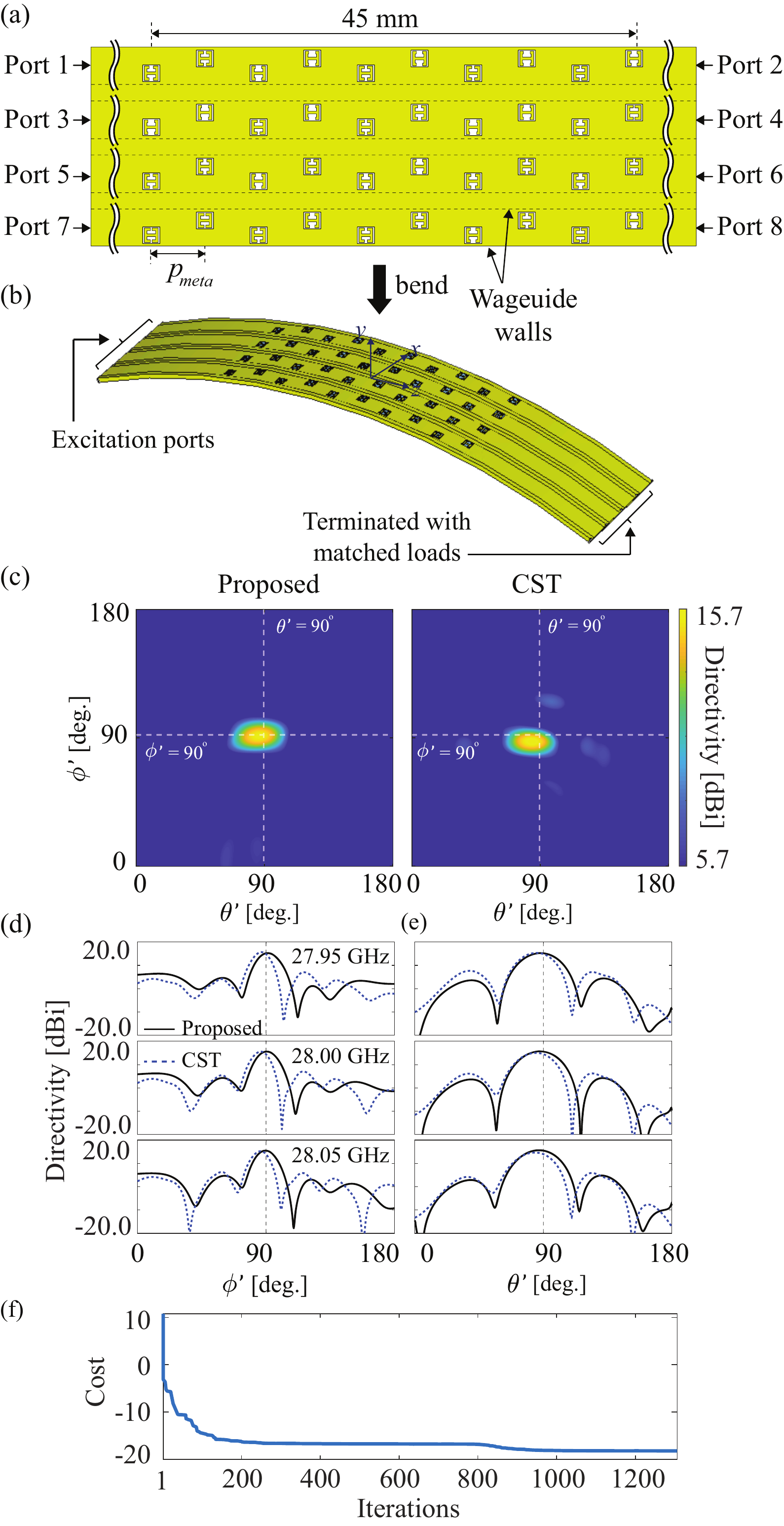}
\caption{(a) Schematic of the designed cylindrical conformal waveguide-fed metasurface antenna before bending (top view). The dotted lines indicate the side walls of each metasurface. (b) Bird-eye view of the designed metasurface antenna wraping a cylindrical surface (with a radius of $r_{bend}$) along the circumferential direction. The arc length is $98$ mm. (c) Directivity pattern of the designed conformal metasurface antenna at 28 GHz. the plane of (d) The cross-sectional plot of radiation patterns for the swept frequencies (at the plane of $\theta^{'}=90^{\circ}$). (e) The evolution of the cost function as a function of the number of iterations.}
\label{Fig4_Ant_Conformal}
\end{figure}

Following the outlined approach, we designed a conformal array of 4 rectangular waveguide-fed metasurfaces, each with 10 metamaterial elements. The schematic of the designed metasurface before bending is depicted in Fig. \ref{Fig4_Ant_Conformal}(a) and after bending is illustrated in Fig. \ref{Fig4_Ant_Conformal}(b). The width and height of a cross-section of the waveguide are $a=3.41$ mm, $b=0.762$ mm. The waveguides are filled with Rogers 4003 circuit board, and their spacing for the unbent structure is $s_{wg}=5.0$ mm. The metamaterial elements are etched into the curved top plates of the waveguides. The radius of curvature is $111.4$ mm, and the arc length is $98.0$ mm. We assume the center of the cylindrical surface of the top plate is located at the origin, as indicated in Fig. \ref{Fig4_Ant_Conformal}(b). For the design, we used the following cost function, given by
\begin{equation} \label{costfun_conformal_array}
\begin{aligned}
\text{cost} = -\Big[\text{min}\Big(D_{dB,odd}\left(f_{op},\mathcal{S},\phi_{tar},\theta_{tar}\Big)\right)  \\ + 0.5\times \text{min}\Big(SLL_{dB,odd}\left(f_{op},\mathcal{S}\right)\Big)\Big],\\
\end{aligned}
\end{equation}
where we used the operating frequencies of $f_{op}=\{27.95\text{ GHz},28.0 \text{ GHz},28.05 \text{ GHz}\}$. $D_{dB,odd}$ represents the directivity when odd-numbered ports are excited by the same excitation power and $90^{\circ}$ phase delay for the port 1 and 7. Such a diversity in phase of the feed wave turned out to be used to suppress unwanted grating lobes \cite{boyarsky2020grating}. In (\ref{costfun_conformal_array}), the target direction was chosen to be $(\phi^{'}_{tar},\theta^{'}_{tar})=(90^{\circ},90^{\circ})$, where the angles are defined as $\phi'=\tan^{-1}(y/z)$ and $\theta'=\tan^{-1}(-\sqrt{y^2+z^2}/x)$. Note that the directivity is used in (\ref{costfun_conformal_array}), which may lead to low efficiency. The estimation of antenna gain of curved metasurface antenna will allow optimization of the gain/efficiency and is left as future work.

Using the setup, the optimal set of gap sizes $\mathcal{S}$ was sought by using the CMA-ES optimizer. In the process, we considered the gap size of individual elements as a tuning parameter that can be controlled to reduce the cost function in (\ref{costfun_conformal_array}) for the operating frequency $f=f_{op}$. The stopping criterion for the optimization process was chosen to be $1\%$ of the relative change in the cost. The criterion can be altered to speed up/down the optimization process; the value of $1\%$ is empirical and turned out to be appropriate for our purpose. Note that the set of polarizabilities found by the optimizer may not lead to the optimal efficiency of the antenna as the cost function in (\ref{costfun_conformal_array}) does not consider the efficiency. Also, metasurface antennas generally require sufficiently many elements to couple most of the energy out to free space \cite{smith2017analysis}, while we have a limited number of elements in each waveguide for full-wave verification.


To verify the proposed design method, we analyzed the designed antenna through the full-wave numerical analysis, where a large perfect conducting, conformal surface with the same radius of curvature is incorporated to suppress the diffraction of the radiated fields at the edges.

Figure \ref{Fig4_Ant_Conformal}(c) shows the predicted and simulated directivity patterns at $28.0$ GHz, demonstrating good agreement. In the predicted pattern, the main beam is formed at $\phi^{'}_{}=90.3^{\circ}, \theta^{'}_{}=88.2^{\circ}$. The main beam of the simulated pattern is at $\phi^{'}_{}=86.8^{\circ}, \theta^{'}_{}=86.8^{\circ}$, which is slightly offset from the target direction. The predicted and simulated directivity values at $28.0$ GHz are $15.7$ dBi and $16.0$ dBi at $28.0$ GH, respectively. SLL of the predicted pattern is computed to be $9.1$ dB and that of the simulated pattern is $7.6$ dB at $28.0$ GHz. To confirm the operating bandwidth of the antenna, we show the cross-sectional plots of the radiation patterns at $\theta^{\prime}=90^{\circ}$ and $\phi^{\prime}=90^{\circ}$ in Figs. \ref{Fig4_Ant_Conformal}(d) and (e). As depicted in Figs. \ref{Fig4_Ant_Conformal}(d) and (e), the shapes of the main beam are maintained over the swept frequencies, and the beams are directed toward the target direction at the swept frequency points, confirming the operation bandwidth of $100$ MHz.

Figure \ref{Fig4_Ant_Conformal}(f) shows the evolution of the cost function, where the stopping criterion was met after $1142$ iterations. As shown in Fig. \ref{Fig4_Ant_Conformal}(f), the cost decreases rapidly in the early stage of the optimization process, and converges to $-18.2$. After 214 iterations (among the total of $1306$), the cost reached $-16.4$, which is $10\%$ higher than that of the converged cost value.

\section{Conclusion}
We have presented a design method for the conformal array of rectangular waveguide-fed metasurface antennas using the coupled dipole method, with modifications on each metamaterial element's location and orientation. In the proposed design method, we combined the analytic model and CMA-ES optimizer to find the optimal set of geometric parameters of the metamaterial elements to achieve desired radiation patterns. While approximate, we showed via numerical studies that the proposed approach can be used as an effective design tool for the cylindrical conformal array of metasurfaces with a large bend radius. The applications of the proposed method includes the design of the conformal metasurface antennas for a variety of applications such as wireless communication systems, security screening, and computational imaging.

\section*{Appendix A \\Near-field coupling of Metamaterial Elements}
\makeatletter
\def\thefigure{A\@arabic\c@figure}
\def\theequation{A\@arabic\c@equation}
\makeatother

\setcounter{equation}{0}
\setcounter{figure}{0}

\begin{figure}[!b]
\centering
\includegraphics[width=3.5in]{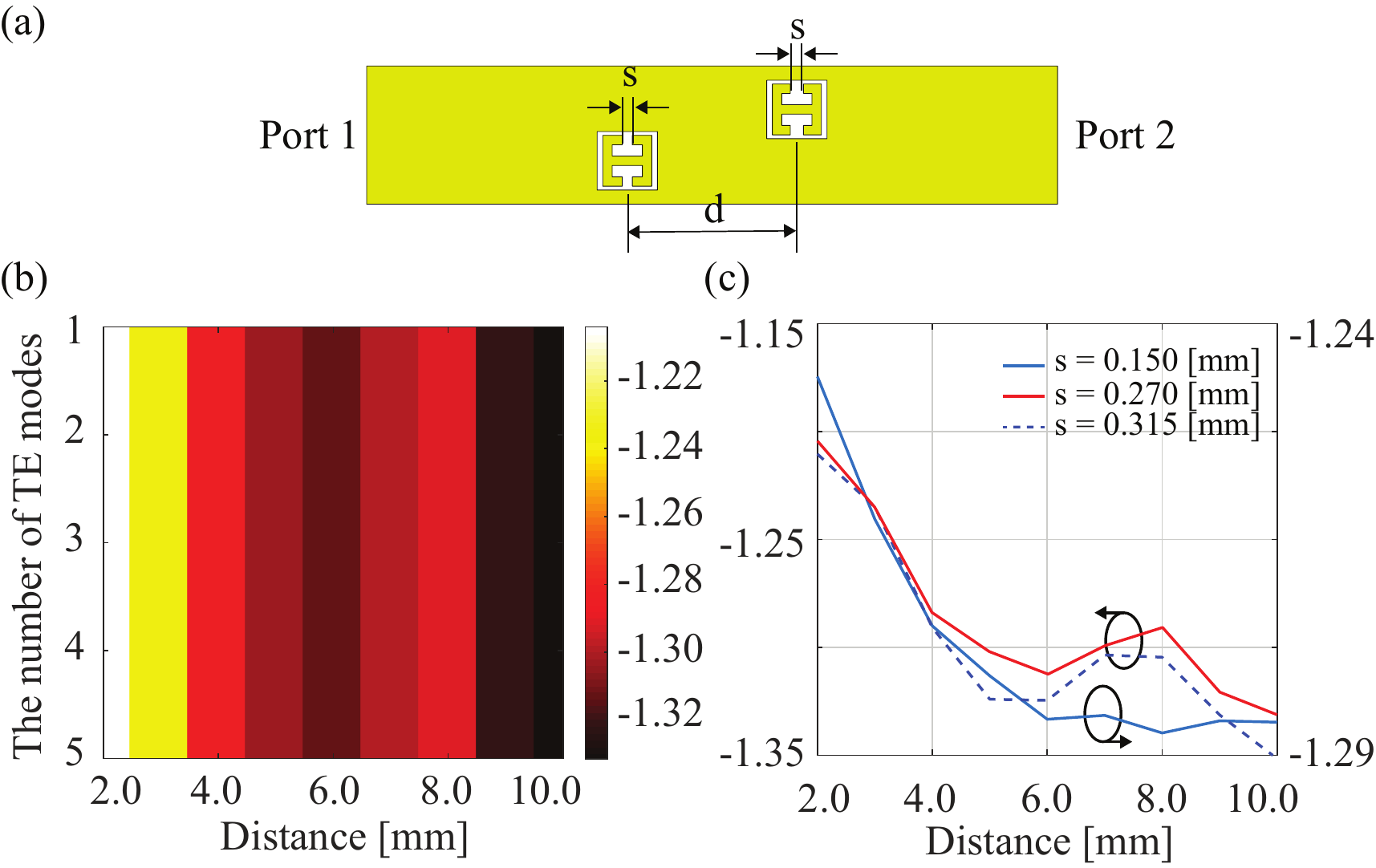}
\caption{(a) Simulation setup for calculating the averaged error in S-parameter. (b) The averaged error in log scale, i.e., $\log_{10}(\Delta_{S})$ for $s=0.27$ mm. (c) The averaged error in log scale for the swept $s$ when the first 5 TE modes are included.}
\label{Fig_AppA1}
\end{figure}

We analyze the effect of the near-field coupling of adjacent metamaterial elements and investigate the minimum allowed distance to ensure the validity of the analytic model in subsection \ref{sec:theory}. Since the proposed analytic model considers the dipolar response of metamaterial elements and their interactions, metamaterial elements need to be placed at distances where the higher-order moments have sufficiently decayed. To analyze the error from ignoring the higher-order moments, we consider a metasurface consisting of the rectangular waveguide in Section \ref{sec:design_ex} and two identical metamaterial elements with a spacing of $d$, depicted in Fig. \ref{Fig_AppA1}(a). To quantify the error, we employ the following measure, expressed as
\begin{equation} \label{sparam_avg_error}
\begin{aligned}
\Delta_{S} = \frac{1}{N_f}\sqrt{\sum_{i=1}^{N_f}\left|S_{11}^{prop}-S_{11}^{CST}\right|^2+\left|S_{21}^{prop}-S_{21}^{CST}\right|^2},
\end{aligned}
\end{equation}
where $N_f = 21$ represents the number of swept frequencies and $f=27-29$ GHz. $S^{prop}_{ij}$ and $S^{CST}_{ij}$ are S-parameters calculated from the proposed model and full-wave simulations, respectively.

Figure \ref{Fig_AppA1}(b) shows the computed averaged error $\Delta_{S}$ for the gap size $s=0.27$ mm. As shown in Fig. \ref{Fig_AppA1}(b), the averaged error decreases as the spacing $d$ increases except for $d=7.0$ mm and $8.0$ mm. It is also shown that the effect of the number of TE modes included in the model on the averaged error is negligibly small for the swept distance $d$, indicating that the contribution of TE$_{10}$ mode to the scattered field by the metamaterial element is dominant. To analyze cases with different gap sizes, we computed the averaged error $\Delta_{S}$ for $s=0.15$ mm, $0.27$ mm, and $0.315$ mm by including the first 5 TE modes in computing the scattered field by the elements. Figure \ref{Fig_AppA1}(c) shows the computed error. As depicted in Fig. \ref{Fig_AppA1}(c), the computed error decreases as the spacing $d$ increases except near $d=7.0$ mm. From these analyses, it may be reasonable to set the minimum allowed distance to be $4.0$ mm ($=\lambda_g/2.58$ at 28.0 GHz).

\section*{Acknowledgment}
This work was supported by the Air Force Office of Scientific Research (AFOSR, Grant No.: FA9550-18-1-0187 and FA9550-18-1-0526).

\bibliographystyle{IEEEtran}
\bibliography{references.bib}

\end{document}